\documentclass[twocolumn]{aastex62}

\shorttitle{Nicolaou et al.}
\shortauthors{Nicolaou et al.}
\begin{document}

\title{The Impact of Turbulent Solar Wind Fluctuations on Solar Orbiter Plasma Proton Measurements}

\correspondingauthor{Georgios Nicolaou}
\email{g.nicolaou@ucl.ac.uk}

\author{G. Nicolaou}
\affiliation{Department of Space and Climate Physics, Mullard Space Science Laboratory, University College London, Dorking, Surrey, RH5 6NT, UK }

\author{D. Verscharen}
\affiliation{Department of Space and Climate Physics, Mullard Space Science Laboratory, University College London, Dorking, Surrey, RH5 6NT, UK }
\affiliation{Space Science Center, University of New Hampshire, NH, USA}

\author{R. T. Wicks}
\affiliation{Department of Space and Climate Physics, Mullard Space Science Laboratory, University College London, Dorking, Surrey, RH5 6NT, UK }

\author{C. J. Owen}
\affiliation{Department of Space and Climate Physics, Mullard Space Science Laboratory, University College London, Dorking, Surrey, RH5 6NT, UK }

%% Note that the \and command from previous versions of AASTeX is now
%% depreciated in this version as it is no longer necessary. AASTeX
%% automatically takes care of all commas and "and"s between authors names.

%% AASTeX 6.2 has the new \collaboration and \nocollaboration commands to
%% provide the collaboration status of a group of authors. These commands
%% can be used either before or after the list of corresponding authors. The
%% argument for \collaboration is the collaboration identifier. Authors are
%% encouraged to surround collaboration identifiers with ()s. The
%% \nocollaboration command takes no argument and exists to indicate that
%% the nearby authors are not part of surrounding collaborations.

%% Mark off the abstract in the ``abstract'' environment.

%%%%%%%%%%%
%%% ABSTRACT %%
%%%%%%%%%%%
\begin{abstract}
Solar Orbiter will observe the Sun and the inner heliosphere to study the connections between solar activity, coronal structure, and the origin of the solar wind. The plasma instruments on board Solar Orbiter will determine the three-dimensional velocity distribution functions of the plasma ions and electrons with high time resolution. The analysis of these distributions will determine the plasma bulk parameters, such as density, velocity, and temperature. This paper examines the effects of short-time-scale plasma variations on particle measurements and the estimated bulk parameters of the plasma. For the purpose of this study, we simulate the expected observations of solar wind protons, taking into account the performance of the Proton-Alpha Sensor (PAS) on board Solar Orbiter. We particularly examine the effects of Alfv\'enic and slow-mode-like fluctuations, commonly observed in the solar wind on timescales of milliseconds to hours, on the observations. We do this by constructing distribution functions from modeled observations and calculate their statistical moments in order to derive plasma bulk parameters. The comparison between the derived parameters with the known input allows us to estimate the expected accuracy of Solar Orbiter proton measurements in the solar wind under typical conditions. We find that the plasma fluctuations due to these turbulence effects have only minor effects on future SWA-PAS observations.
\end{abstract}
%%%%%%%%%%%
%%%%%%%%%%%

%% Keywords should appear after the \end{abstract} command.
%% See the online documentation for the full list of available subject
%% keywords and the rules for their use.

%% From the front matter, we move on to the body of the paper.
%% Sections are demarcated by \section and \subsection, respectively.
%% Observe the use of the LaTeX \label
%% command after the \subsection to give a symbolic KEY to the
%% subsection for cross-referencing in a \ref command.
%% You can use LaTeX's \ref and \label commands to keep track of
%% cross-references to sections, equations, tables, and figures.
%% That way, if you change the order of any elements, LaTeX will
%% automatically renumber them.
%%
%% We recommend that authors also use the natbib \citep
%% and \citet commands to identify citations.  The citations are
%% tied to the reference list via symbolic KEYs. The KEY corresponds
%% to the KEY in the \bibitem in the reference list below.

%%%%%%%%%%%%%%%
%%%  INTRODUCTION  %%%
%%%%%%%%%%%%%%%
\section{Introduction} \label{sec:intro}

\added{As the solar wind expands into the heliosphere, it develops a strong turbulent character \citep[e.g.,][]{Tu1995, Marsch2006, Bruno2013}},\deleted{The solar wind plasma is highly dynamic} with spatial and temporal variations over a wide range of scales \citep[e.g.,][]{Goldstein1995,Verscharen2019}. \added{Numerous studies have revealed the nature of the turbulence at different scales, identifying Alfv\'{e}nic fluctuations \citep{Belcher1971}, magnetoacoustic (fast and slow MHD) fluctuations and pressure-balanced structures \citep{Tu1995,Bruno2013} at large scales, and the contribution of fluctuations with polarization properties of kinetic Alfv\'{e}n waves, slow modes, and whistler modes at small scales \citep{Gary2009}. The presence of these fluctuations makes the study of the plasma kinetic state at a given time challenging, as the plasma kinetic state constantly changes self-consistently in response to the turbulence fluctuations \citep[e.g.,][\& references therein]{Marsch2006}. Moreover, at small scales, the plasma and field fluctuations do not follow Gaussian statistics and exhibit properties of intermittency, increasing the complexity of the system \citep[e.g.,][]{Matthaeus2015, Wan2016}.}

In-situ plasma observations provide the information to study the kinetic properties and the dynamics of the solar wind. The three-dimensional (3D) velocity distribution function (VDF) of the plasma particles, at a given time, contains the information to derive the plasma bulk parameters, such as the density, velocity, and temperature. Past and future solar wind missions have been designed to study the solar wind by obtaining the 3D VDFs of its component populations with a time resolution ranging from a few seconds to more than 1 minute. However, the effect of the highly-dynamic nature of the solar wind on the accuracy of the measurements has not been often considered.\par

For example, the Helios probes were launched in the mid 1970s and operated in a heliocentric orbit, reaching a perihelion of about 0.3$~\mathrm{au}$ to study the solar wind in the inner heliosphere for the first time. The plasma experiment E1 on board Helios was designed to measure the solar wind plasma particles and determine their 3D VDFs \citep{Schwenn1975, Rosenbauer1977}. In the experiment's nominal operation mode, Helios data provided the full 3D VDF of protons every $\sim40~\mathrm{s}$. \par

The Wind spacecraft was launched in 1994 and is dedicated to investigate basic plasma processes in near-Earth space. It has been in a halo orbit around $L_1$ since 2004. Wind's Solar Wind Experiment (SWE) is a comprehensive plasma instrument, measuring the distributions of protons and heavier ions \citep{Ogilvie1995}. It carries a Faraday cup subsystem which, in a nominal mode, provides the measurements to determine the densities, bulk velocities, and temperatures of solar wind ions every 92$~\mathrm{s}$. Wind's three-dimensional plasma and energetic particle investigation instrument, Wind/3DP  \citep{Lin1995}, carries a set of proton electrostatic analyzers (PESA) and a set of electron electrostatic analyzers (EESA) which measure the 3D VDFs of the corresponding species every 3$~\mathrm{s}$.\par

Solar Orbiter is scheduled for launch in February of 2020. It is designed to study the inner heliosphere, which in part it will do by measuring the solar wind plasma in-situ with a higher time resolution than previous missions. The Solar Wind Analyser's Proton - Alpha Sensor (SWA-PAS) on board Solar Orbiter, is an electrostatic analyzer that will measure the 3D VDF every 4$~\mathrm{s}$.\par

There are technological limitations that prevent simultaneous observations of the entire 3D VDF in infinitesimal time intervals. Typical plasma sensors, such as those mentioned above, scan through energy and flow direction of the particles in discrete consecutive steps, measuring the particle flux at each step in a given time interval (acquisition time). As a result, within the measurement time for a full 3D VDF, the individual instrument samples are affected by any fluctuations of the distribution function that occur on shorter time scales. Such small-scale variations affect the observed VDF and thus the estimated plasma bulk parameters. For example, when a relatively sharp discontinuity passes over the spacecraft, while the instrument performs a 3D VDF scan, the bulk velocity may rapidly change. In such a case, the instrument may observe parts of two very different VDFs for each `half' of its scan. If the resulting observation is interpreted as they were one VDF, the results are distorted. Any later analysis of moments will be wrong, as they will neither correspond to the upstream nor the downstream plasma region, nor indeed any part of the boundary itself. \par

In an example, \citet{Verscharen2011} show that wave activity can lead to artificial temperature anisotropies in the observed plasma distributions. Large-amplitude waves can shift the VDF in the direction perpendicular to the background magnetic field. Since these fluctuations occur at time scales smaller than the instrument's sampling time, the observed average distribution exhibits a broadening in the perpendicular direction, which eventually could be misinterpreted as an intrinsic temperature anisotropy. In a more recent study, \citet{Nicolaou2015a} demonstrate that random variations in the plasma bulk parameters result in broader VDFs, which eventually lead to a bias towards higher temperatures. The authors consider observations of plasma ions in the distant Jovian tail by the Solar Wind Around Pluto Instrument \citep[SWAP;][]{McComas2008} on board New Horizons.\par

In this paper, we predict the effects of temporal variations due to turbulence on measurements with Solar Orbiter's SWA-PAS. \added{We adopt the well-established forward-modeling technique by modeling the instrument response in a simulated plasma environment \citep[see also][]{Vaivads2016, Cara2017, Wilson2017, Kim2019}}. We specifically consider the characteristic solar wind  plasma behavior due to Alfv\'enic and slow-mode-like waves turbulence. Early observations of the solar wind \citep[e.g.,][]{Belcher1971} showed that proton velocity and magnetic field fluctuations are highly correlated for a majority of the time. This is the characteristic signature of Alfv\'en waves; plasma waves with fluctuations transverse to the magnetic field direction. Detailed analysis over the past 40 years has shown that Alfv\'enic modes carry the majority of the energy in the free-flowing solar wind \citep[e.g.,][]{Roberts2010,Wicks2013}. More recent statistical analyses have shown that there is a minor component of slow-mode waves \citep[e.g.,][]{Klein2012,Verscharen2017} which are longitudinal compressive waves. These two wave fields act to distort the proton VDF measurement by fluctuating the plasma on the time scale over which the observation is made \citep{Verscharen2011}.  

In this study, we model the expected observations in such turbulent conditions, and quantify the error of the plasma parameters derived from the moments of the 3D VDF. Our study could be extended for the diagnosis of the errors of SWA-PAS plasma observations. In the following section, we describe SWA-PAS, and, in Section \ref{sec:meth}, we describe the method we use to simulate the expected observations and our standard techniques to analyze them. In Section \ref{sec:res}, we present our results, which we discuss in detail in Section \ref{sec:dis}. We also discuss and compare the expected errors in the measurements of previous missions. The model that we use for the solar wind turbulence is included in the Appendix.

%%%%%%%%%%%%%%%%%%
%%% INSTRUMENTATION  %%%%%
%%%%%%%%%%%%%%%%%%
\section{Instrumentation} \label{sec:ins}

SWA consists of three sensors: i) The Proton-Alpha Sensor (SWA-PAS), ii) the Electron Analyser System (SWA-EAS), and iii) the Heavy Ion Sensor (SWA-HIS). The three sensors share a common Data Processing Unit (DPU) and are designed to measure the 3D VDFs of the solar wind particles. We use an idealized model of SWA-PAS, which is designed to observe the energy-per-charge range from 0.2 to 20$~\mathrm{keV/q}$. We consider a specific operation mode in which this range is covered in 96 exponentially spaced steps with a resolution of $\Delta E/E \sim7.5 \%$. The azimuth field of view (F.O.V.) ranges from $-24^{\circ}$ to + $42^{\circ}$ with respect to the Sun direction, accounting for the expected range of the aberration angle, and is covered by 11 sectors that consist of individual channel electron multipliers (CEMs). The elevation F.O.V. ranges from $-22.5^{\circ}$ to + $22.5^{\circ}$ with respect to the Sun direction and is covered by 9 electrostatic steps performed by the electrostatic deflector system (see Figures \ref{fig:pas} a and b). In the operation mode we consider here, the instrument performs one full 3D scan by repeating 9 elevation scans for each of the 96 energy steps, while for each energy and elevation pair, the 11 CEMs record the azimuth directions simultaneously. The instrument scans energies from highest to lowest, while it scans the elevation angles from top to bottom and from bottom to top, in a consecutive order (see Figure \ref{fig:pas} c). The acquisition time ($\Delta \tau$) for each energy and elevation direction is $\sim$1~ms. A full 3D VDF is obtained in $\sim$1~s, followed by $\sim$3~s of no measurement, resulting in an overall $\sim$4~s cadence. We develop a model of SWA-PAS based on its initial calibration and ideal response for simplicity. We also neglect the voltage transition time during the energy-elevation scans.

\begin{figure*}[ht!]  \label{fig:pas}
\centering
\scalebox{1.2}{\plotone{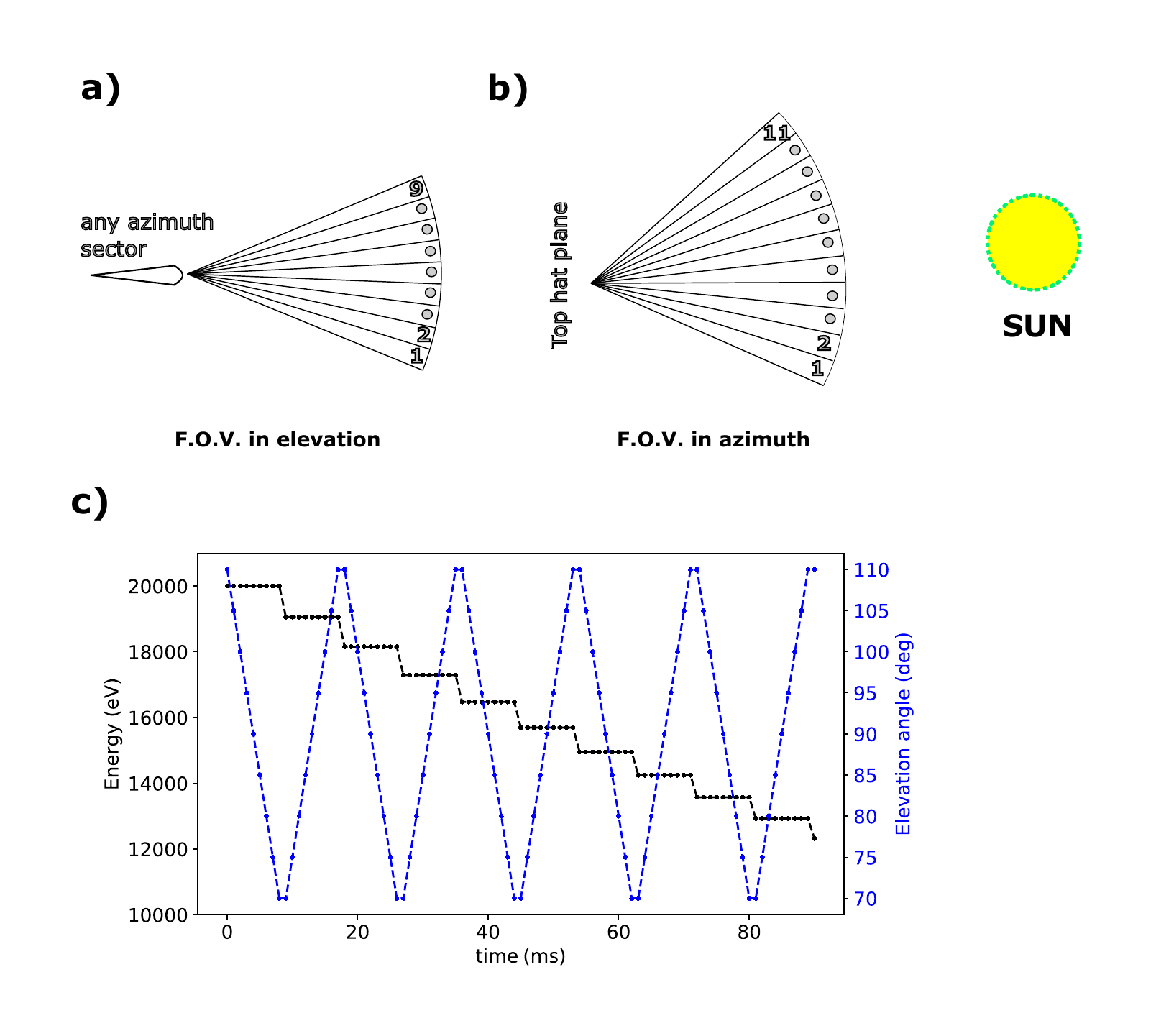}}
\caption{a) The elevation and the b) azimuth field of view of our SWA-PAS model. c) an example of the instrument's energy $\times$ elevation scans during modeled operations. We show the first ten energy $\times$ elevation scans, as the instrument scans the elevation directions from top to bottom and from bottom to top in a consecutive order (blue line) for each energy step, starting from the highest energy (black line).}
\label{fig:pas}
\end{figure*}

%%%%%%%%%%%%%%%%%%%%%%%%
%%%  Data and Instrument Simulation  %%%%%%
%%%%%%%%%%%%%%%%%%%%%%%%
\section{Data and Instrument Simulation} \label{sec:meth}

SWA-PAS will measure the plasma at heliospheric distances between $\sim$0.3 and $\sim$1 au. Within this range, the average density $n_0$ is expected between $\sim$1 and $\sim$50 cm$^{-3}$, the average temperature $T_0$ between few eVs and $\sim$50 eV, the average magnetic field $B_0$ between $\sim$1 and $\sim$40 nT, and the average bulk speed $\sim$500 km\,s$^{-1}$ \citep[e.g,][]{Barouch1977,Freeman1988}. For this paper, we model plasma turbulence for $n_0$ = 20 cm$^{-3}$, $T_0$ = 20 eV, $B_0$ = 10 nT, and $u_0$ = 500 km\,s$^{-1}$, which we consider typical values within the expected ranges. For these background plasma parameters and magnetic field, the Alfv\'{e}n speed $V_{\mathrm{A}} \, \sim50 \, \mathrm{km \, s^{-1}}$, the plasma beta $\beta_{\mathrm{p}}\,\sim \mathrm{1.6}$, and the proton gyroradius $\rho_{\mathrm{g}}\, \sim \mathrm{65\, km}$. We model the fluctuations of the plasma parameters considering Alfv\'{e}nic and slow-mode-like turbulence. The Alfv\'{e}nic component introduces fluctuations mainly in the velocity component perpendicular to the magnetic field. The slow-mode-like component is the minor component of the turbulent spectrum, but introduces density fluctuations in the frequency domain above the kinetic scales. \added{For simplicity, we construct the turbulence spectrum considering that the waves are frozen in the plasma flow, which is known as the Taylor's hypothesis \citep{Taylor1938}. This simplification allows us to model a turbulent spectrum with power levels and polarization properties that match typical spacecraft observations. However, it is currently a matter of ongoing research to what degree Taylor's hypothesis is applicable to the types of fluctuations we discuss \citep[e.g.,][]{Howes2014, Klein2014, Klein2015,Perri2017,Narita2017,Bourouaine2018}.} We describe our calculation of the solar wind input distributions and their fluctuations due to turbulence in the Appendix. In the next subsections, we define our instrument model and the analysis of modeled measurements for specific 3D solar wind input VDFs, taking into account the SWA-PAS response.   

           %%%%%%%%%%%%%%%
           %%%  METH: PSEUDO    %%%
           %%%%%%%%%%%%%%%
\subsection{SWA-PAS Observation Model} \label {subsec:pseudo}

SWA-PAS measures the number of particles that enter the instrument aperture in each acquisition step $\tau$ at the specific energy $E(\tau)$, elevation $\Theta(\tau)$ and azimuth sector $\Phi$. The measured energy and elevation directions are functions of time, based on the sequential sampling process of the sensor (see Section \ref{sec:ins}). We calculate the expected counts $C(E(\tau),\Theta(\tau),\Phi,\tau)$ to be obtained at each acquisition step $\tau$ based on our modeled distribution $f$ as

\begin{eqnarray} \label{eq:counts}
C(&E&(\tau),\Theta(\tau),\Phi,\tau) = \nonumber \frac{2}{m^2} \int \limits_{E-\frac{\Delta E}{2}}^{E+\frac{\Delta E}{2}} \int \limits_{\Theta - \frac{\Delta \Theta}{2}}^{\Theta+\frac{\Delta \Theta}{2}} \int \limits_{\Phi - \frac{\Delta \Phi}{2}}^{\Phi + \frac{\Delta \Phi}{2}}\int \limits_{\tau-\frac{\Delta \tau}{2}}^{\tau+\frac{\Delta \tau}{2}} \nonumber \\
&&A_{\mathrm{eff}}(\epsilon,\theta,\phi)f(\epsilon,\theta,\phi,t) \epsilon \, \mathrm{d} \epsilon \, \mathrm{\replaced{sin}{cos}} \theta \mathrm{d} \theta \, \mathrm{d} \phi \, \mathrm{d}t,
\end{eqnarray}

\noindent where $m$ is the mass of a measured particle and $A_{\mathrm{eff}}$ is the effective area of the sensor. The 3D VDF $f$ is expressed in spherical coordinates, where $\epsilon$ is the particle energy, $\theta$ the elevation angle, $\phi$ the azimuth angle, and $t$ the time. The energy resolution $\Delta E/E$ and the angular resolution in elevation and azimuth direction, $\Delta \Theta$ and $\Delta \Phi$ respectively, are considered constant for simplicity. As in \citet{Nicolaou2018}, we assume that $A_{\mathrm{eff}}$ is a discrete function of the elevation step $\Theta$ only, i.e. $A_{\mathrm{eff}}(\epsilon,\theta,\phi) \equiv A_{\mathrm{eff}}(\Theta)=A_{0}/\replaced{\sin}{\cos}\Theta $. The independence of $A_{\mathrm{eff}}$ on $\Phi$ assumes that the detection efficiency of the 11 CEMs in PAS is identical. Additionally, since we want to investigate specifically the effects of short period turbulence fluctuations on the expected observations, we intentionally exclude statistical uncertainties (Poisson error) and any other physical source of statistical and systematical errors; such as background radiation, electronics noise, and contamination of the detectors. With these simplifications, we calculate the expected counts as

\begin{eqnarray} \label{eq:counts_simp1}
C(&E&(\tau),\Theta(\tau),\Phi,\tau)=\frac{2}{m^2} A_{0}E^2 \frac{\Delta E}{E} \, \Delta \Theta \, \Delta \Phi \nonumber \\
&&\times \sum_{\tau-\frac{1}{2}\Delta\tau}^{\tau+\frac{1}{2}\Delta\tau}{f(\epsilon=E,\theta=\Theta,\phi=\Phi,t)\, \mathrm{d}t},
\end{eqnarray}
\noindent in which the integral over time in Equation (\ref{eq:counts}) is solved numerically.   

           %%%%%%%%%%%%%%%%%%%
           %%%  METH: PSEUDO analysis    %%%
           %%%%%%%%%%%%%%%%%%%
\subsection{Analysis of SWA-PAS Modeled Observations} \label{subsec:data_analysis}
Most space-plasma analyses assume that $f$ remains constant during a full VDF scan period of the particle instrument. Under this assumption, Equation (\ref{eq:counts}) becomes:

\begin{eqnarray} \label{eq:counts_simp2}
C(&E&,\Theta,\Phi)=\frac{2}{m^2} A_{0} E^2 \frac{\Delta E}{E} \, \Delta \Theta \, \Delta \Phi \nonumber \\
&&\times f(\epsilon=E,\theta=\Theta,\phi=\Phi) \Delta \tau,
\end{eqnarray}

\noindent which we invert to calculate the distribution function from the observed counts as

\begin{equation} \label{eq:f_from_c}
f_{\mathrm{out}}(E,\Theta,\Phi) \approx \frac{m^{2}C(E,\Theta,\Phi)}{2E^2G},
\end{equation}
\noindent where 

\begin{equation} \label{eq:G_factor}
G= A_{0} \frac{\Delta E}{E} \, \Delta \Theta \, \Delta \Phi \, \Delta \tau
\end{equation}

\noindent is the geometric factor of the instrument \citep[for more details, see][]{Nicolaou2018}. 
The common application of Equation (\ref{eq:f_from_c}) in space-plasma analyses introduces inaccuracies if there are changes in $f$ at time scales shorter than the sampling time for a full 3D VDF. \par

In order to construct our modeled observations in a time-varying plasma, we take into account variations during the scanning sequences of the instrument. As the instrument scans in energy and elevation we vary $f$ using a model of Alfv\'enic and slow-mode-like turbulence, suitable for the solar wind (see Appendix \ref{sec:meth_turb}). The turbulent fluctuations cause $f$ to vary in time and so introducing inaccuracies in the determination of $f_{\mathrm{out}}$ from the above assumption of time invariance, as discussed, in Equation \ref{eq:counts_simp2}. We then derive the distribution function from counts using Equation (\ref{eq:f_from_c}) under the discussed assumptions, and calculate its bulk parameters as moments. We compare the derived moments with those used to model the solar-wind plasma in the first place. This comparison allows us to quantify the error of the estimated plasma parameters due to under-resolved variations of the plasma.

%%%%%%%%%%%%%%%%%%%
%%%  RESULTS  %%%%%%%%%%
%%%%%%%%%%%%%%%%%%%
\section{Results} \label{sec:res}

           %%%%%%%%%%%%%%%%
           %%%  RES: TURBULENCE   %%%
           %%%%%%%%%%%%%%%%

\begin{figure*}[ht!] 
\scalebox{1.2}{\plotone{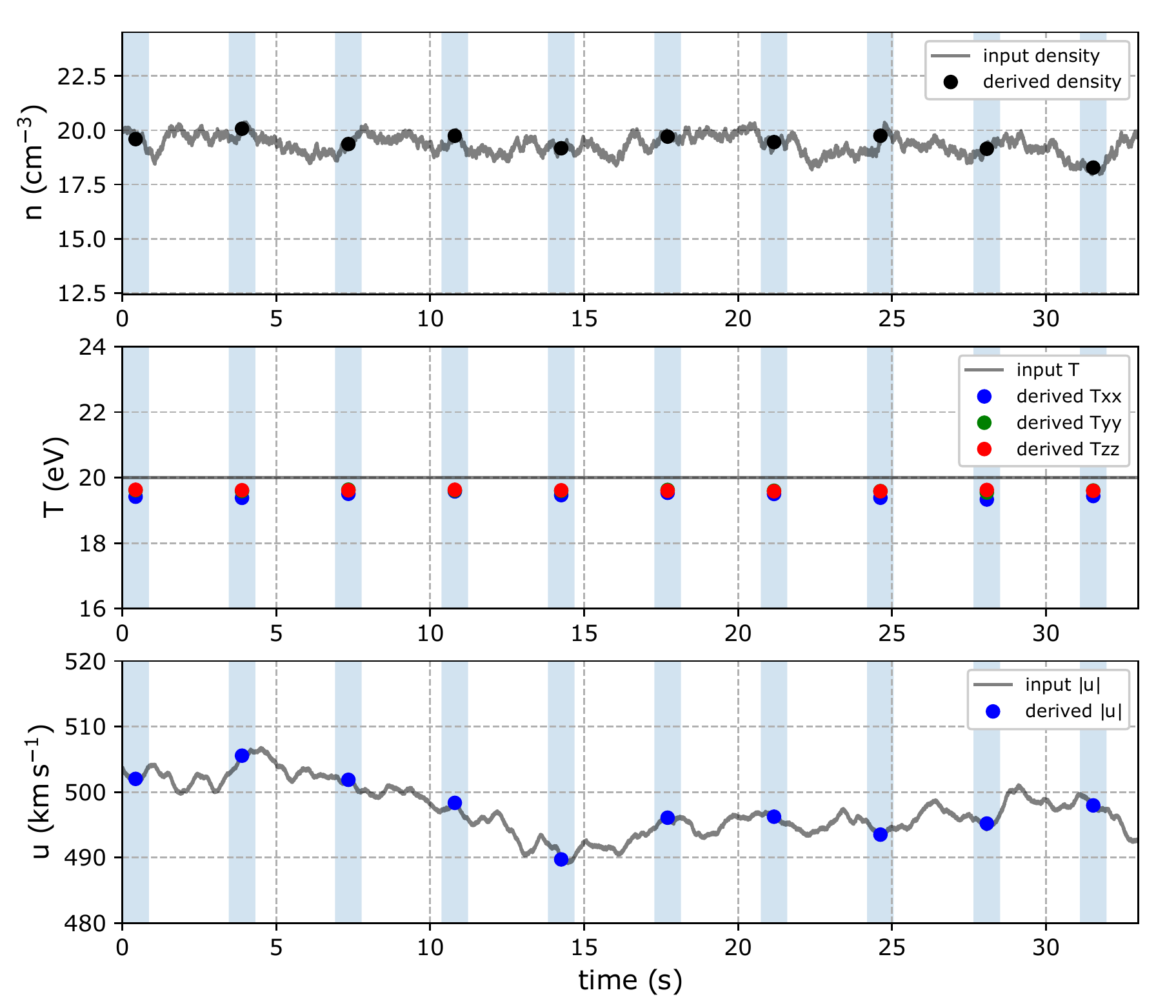}}
\caption{Time series of modeled solar wind with a turbulent spectrum consisting of Alfv\'en waves and slow modes and a comparison to derived moment parameters from the expected SWA-PAS observations at lower resolution. Each panel shows the input data (gray line) and the moments derived from the modeled observations (bullets). The shadowed areas represent the time intervals in which the instrument collects counts to construct an entire 3D VDF. The top panel shows the plasma density ($n=n_0+\Delta n$), the middle panel shows the diagonal elements of the plasma temperature tensor ($T=T_0$), and the bottom panel shows the plasma bulk speed ($u=|\vec{u}_0+\Delta \vec{u}|$). Besides the small systematic \replaced{over}{under}estimation of the plasma density and \deleted{underestimation of the} plasma temperature, the derived moments suggest that the accuracy of SWA-PAS measurements, under typical turbulent solar wind conditions, is remarkably high.}
\label{fig:fluct_obs_mom}
\end{figure*}

\begin{figure*}[ht!] 
\scalebox{1.2}{\plotone{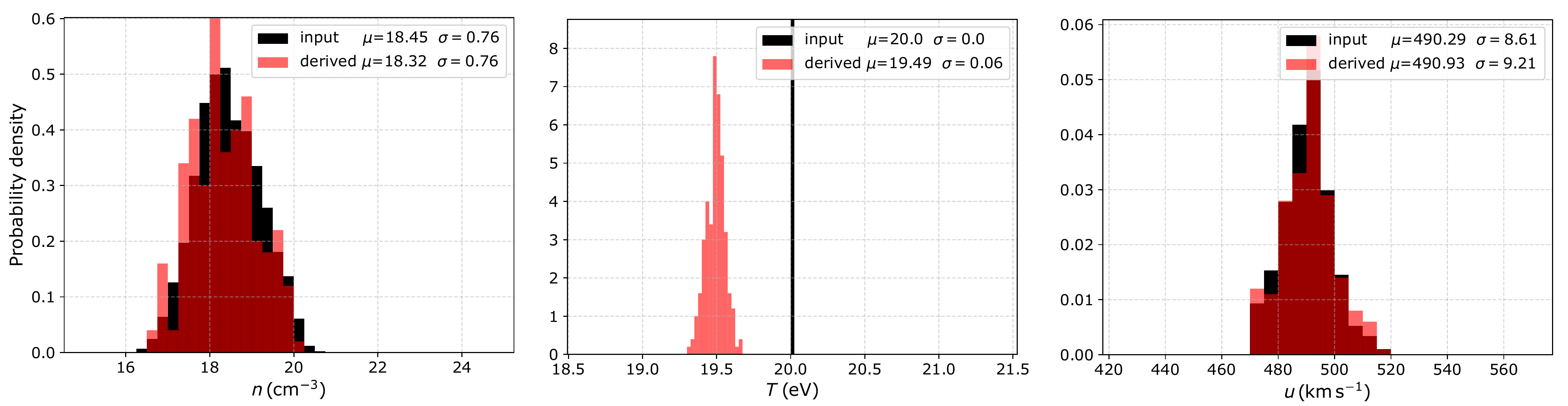}}
\caption{Histograms of the derived moments (red) and the average input moments (black) over the instrument's 3D VDF measurement intervals (black). The left panel shows the plasma density, the middle panel shows the scalar temperature, and the right panel shows the bulk speed. For this analysis, we analyze a sample of 200 modeled observations.}
\label{fig:histograms}
\end{figure*}

Figure \ref{fig:fluct_obs_mom} shows the first 33~s of the modeled solar-wind proton bulk parameters for the input turbulence conditions described in Appendix \ref{sec:meth_turb} and the corresponding analysis of SWA-PAS modeled observations. The derived parameters are, by eye, in good agreement with the input parameters. In order to quantify the error of the estimated parameters due to the modeled turbulence, we construct histograms of density, temperature, and bulk speed as derived from the analysis of 200 observations sampled at random time intervals in our modeled turbulence. These are represented by the \replaced{black}{red} histograms in Figure \ref{fig:histograms}. Overlaid in each panel, we also show a histogram of the mean values of the corresponding input moments over each of the 200 observations (in \replaced{red}{black}). These are the time averages of the input plasma moments, over each of the 200 full 3D instrument scans (approximately 1~s each). Besides small systematic errors associated with the numerical calculation of moments (see also the related discussion in the next section), the difference between the standard deviations of the derived and the input parameters indicates that the error of the derived parameters due to turbulence is remarkably small. Note again that the statistical error of the derived plasma parameters presented here is due to turbulence only, as we do not include any other source of statistical error in our model.

\begin{figure*}[h] 
\scalebox{1.2}{\plotone{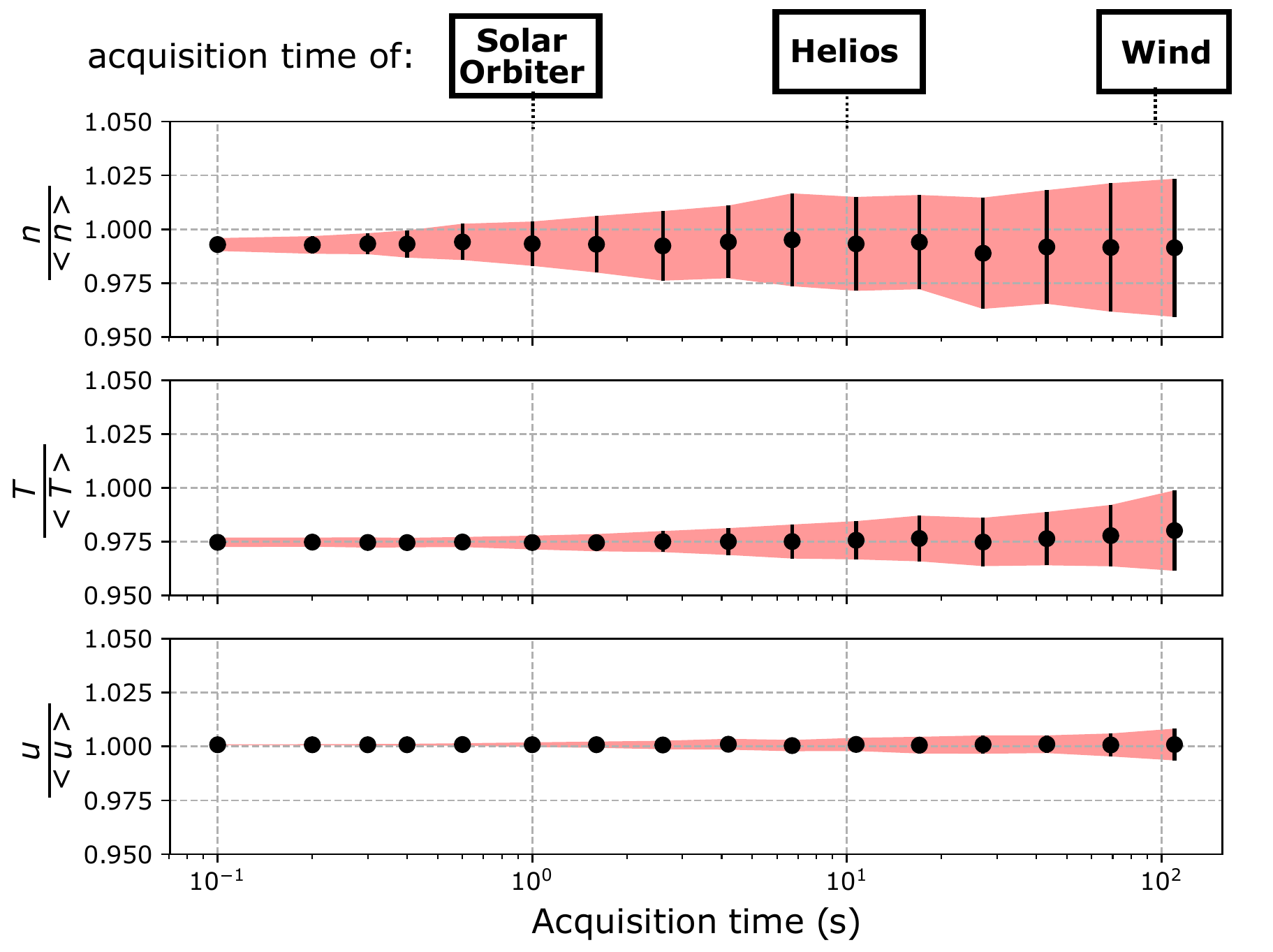}}
\caption{Average values and standard deviations of the normalized derived density (top), scalar temperature (middle) and bulk speed (bottom) for different 3D VDF acquisition times. The values are normalized to their averaged input values over each time interval for a full 3D VDF measurement. The average plasma density is \replaced{over}{under}estimated by \replaced{$\sim$}{less than }1$\%$, the average plasma temperature is underestimated by $\sim$\replaced{3}{2.5}$\%$ and increases slightly with acquisition time, while the average plasma speed is practically calculated with no error. The measured moments exhibit an increasing standard deviation, shown as the red area, as the acquisition time increases. The standard deviation of the normalized density increases from $\sim$1$\%$ to 3.5$\%$ as the acquisition time increases from 1~s to $\sim$100~s. Within the same range of acquisition times, the standard deviation of the normalized plasma temperature increases from $<1\%$ to 2$\%$. The stantard deviation of the normalized derived speed is $<1\%$ for the acquisition times we examine here. For comparison, we indicate the acquisition times of specific missions on the top of the plot.}
\label{fig:derN_derT_derU_vs_acqt}
\end{figure*}

For comparison, we now study the effect of turbulence on measurements taken with different acquisition times. We specifically examine 3D VDF acquisition times ranging from 0.1~s to $\sim$100~s. For each acquisition time, we construct 200 modeled observations recorded at random time intervals in our model turbulence. We normalize the derived density, temperature, and bulk speed of each observation to their average values of the corresponding input moments within the specific acquisition. In Figure \ref{fig:derN_derT_derU_vs_acqt}, we show the mean values (dots) and the standard deviations (red area) of the normalized derived plasma parameters as functions of the acquisition time. As the acquisition time increases, the uncertainty of the moments increases. The plasma density shows the greatest deviation while the plasma speed shows the smallest deviation. In addition, the derived plasma temperature slightly increases with acquisition time, as expected from the analyses by \citet{Verscharen2011} and \citet{Nicolaou2015a}. Nevertheless, even for the highest acquisition time shown, the standard deviation of the derived parameters lies within a few percent of the corresponding average value. In the same figure, we also note the acquisition time of previous missions.

%%%%%%%%%%%%%%%%%%
%%% DISCUSSION  %%%%%%%%
%%%%%%%%%%%%%%%%%%

\section{Discussion and Conclusions} \label{sec:dis}

Our analysis suggests that typical plasma fluctuations due to solar wind turbulence have only minor effects on upcoming SWA-PAS observations. Figure \ref{fig:fluct_obs_mom} demonstrates that the expected measured plasma density and temperature are slightly affected by a realistic turbulence spectrum, while the effects on the estimation of the bulk speed are negligible. The histograms of the derived plasma parameters in Figure \ref{fig:histograms} indicate a small deviation from the corresponding input parameters. Even though the the plasma temperature input is constant with time at 20 eV in our model, because the Alfv\'en wave and slow-mode models used are isothermal, the derived temperature has a standard deviation of $\sim$\replaced{0.07}{0.06} eV. \par

Our comparative study (Figure \ref{fig:derN_derT_derU_vs_acqt}) shows that the plasma turbulence affects less the accuracy of SWA-PAS than it affects the accuracy of previous missions, measuring plasma protons in lower time resolution. The standard deviation of the normalized derived density is $\sim$1\% for the acquisition time of Solar Orbiter, while is $\sim$2\% for the acquisition time of Helios, and 3.5\% for the acquisition time of Wind. The standard deviation of the normalized derived temperature is $<$1\% for the acquisition time of Solar Orbiter, $\sim$1\% for the acquisition time of Helios, and 2\% for the acquisition time of Wind. The standard deviation of the normalized derived speed is $<$1\% for the range of acquisition times we examine here. \par

\begin{figure*}[ht!] 
\scalebox{1.2}{\plotone{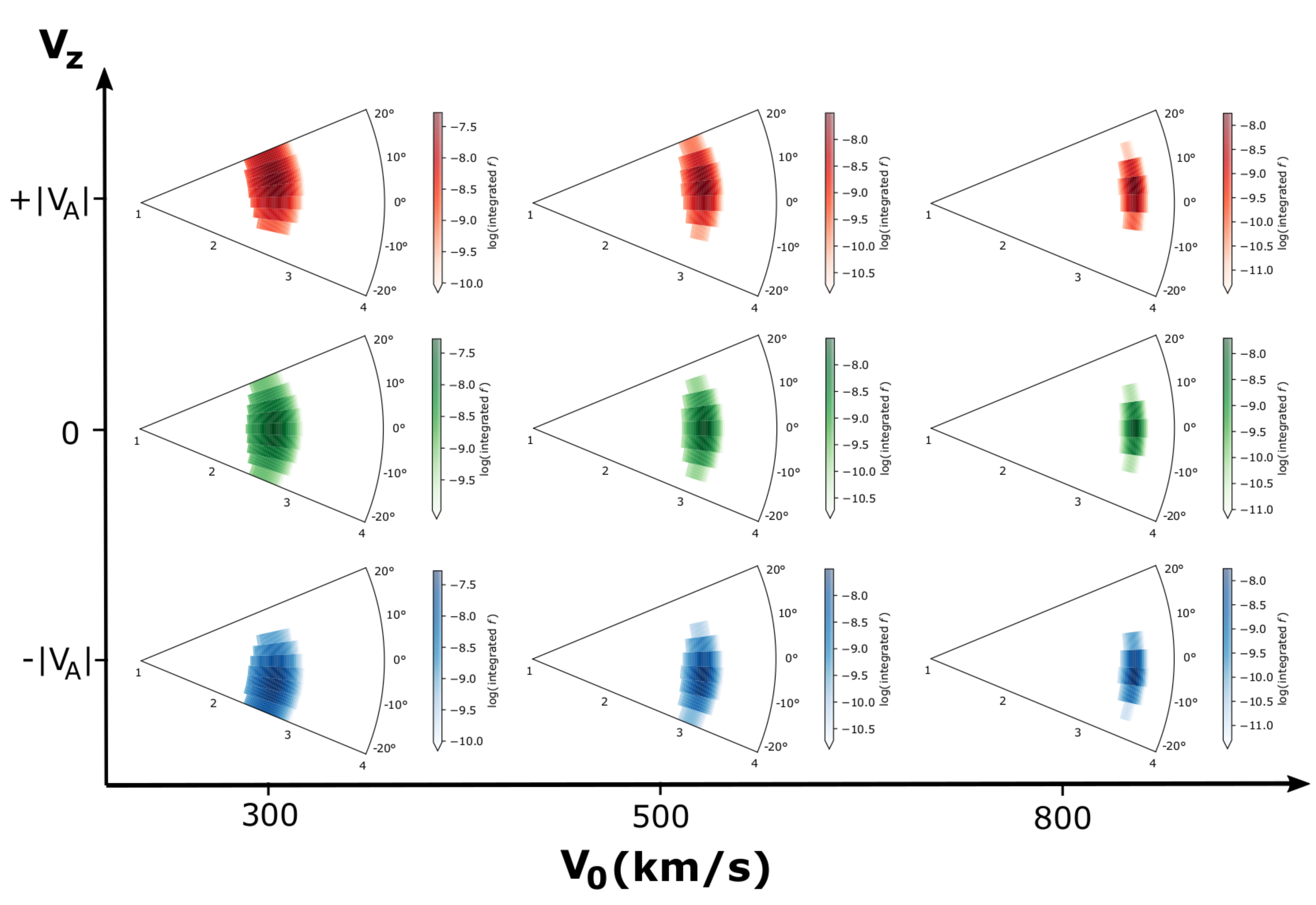}}
\caption{\added{Plasma distribution functions in the instrument frame (integrated over azimuth direction). We assume a plasma with $n_{\mathrm{0}}$ = 20 cm$^{-3}$, $T_{\mathrm{0}}$ = 20 eV and three different background bulk velocities: (left) $u_{\mathrm{0}}$= 300 km s$^{-1}$, (middle) 500 km s$^{-1}$, and (right) 800 km s$^{-1}$. For all the examples, $V_{\mathrm{A}}$ = 50 km$\,s^{-1}$ and for each $u_{\mathrm{0}}$ we consider the cases (bottom) $\vec{u}=u_{\mathrm{0}}\hat{x}-V_{\mathrm{A}}\hat{z}$, (middle) $\vec{u}=u_{\mathrm{0}}\hat{x}$, and (top) $\vec{u}=u_{\mathrm{0}}\hat{x}+V_{\mathrm{A}}\hat{z}$. The angular deviation of the center of the distribution increases with decreasing $u_{\mathrm{0}}$, so that a significant portion of the distribution can leave the F.O.V. if $V_{\mathrm{A}}/u_{\mathrm{0}}$ is large.}}
\label{fig:Alfven_Bulk}
\end{figure*}

Figures \ref{fig:fluct_obs_mom}, \ref{fig:histograms}, and \ref{fig:derN_derT_derU_vs_acqt} show that the plasma density \added{and temperature} \replaced{is}{are} slightly \replaced{over}{under}estimated \added{by $<$1 \% and $\sim$2.5 \% respectively}\deleted{(by $\sim$1$\%$) while the plasma temperature is slightly underestimated (by $\sim$3$\%$)}. Although we intentionally \replaced{did}{do} not include  any source of error in our model, calculating the moments of a distribution function by integrating it in discrete steps, introduces such systematic errors. \added{This error occurs due to the instrument's finite and discrete angular and energy resolution (see also Figure \ref{fig:Alfven_Bulk}). Such an error depends on the plasma parameters, which we will address and correct for in the future, when a complete error analysis of SWA-PAS is available. \par} 
In addition, due to limited efficiency, the instrument cannot resolve the \added{full} tails of the distribution function \added{characterized by particle fluxes that are too low to produce detectable signal}. \deleted{Therefore, its moments (especially those of higher order) are underestimated}\added{As a result, the undetected particles do not contribute to the mathematical calculation of the moments, resulting in an underestimation of the plasma density and temperature \citep[e.g.,][]{Nicolaou2016, Nicolaou2018}. We demonstrate this effect in Figure \ref{fig:n_and_T_vs_Gfactor}, where we plot the derived density and temperature as functions of the instrument efficiency, considering a non-fluctuating plasma with the same background parameters as in our turbulent solar wind model (Section \ref{sec:meth}). We scale the model instrument's geometric factor $G$ by an efficiency multiplier $A$ (i.e., $G\rightarrow AG$) for each synthetic sample. For the value used in this work (i.e., $A=1$, red dashed in Figure \ref{fig:n_and_T_vs_Gfactor}) the error is similar to our results presented in Section \ref{sec:res}. Moreover, the calculated $n$ and $T$ exhibit an asymptotic behavior, approaching the corresponding input values as $A$ increases. We note that in this work we consider constant (with energy and look direction) efficiency (see also Section \ref{subsec:data_analysis}), while the actual instrument efficiency may vary by 40\%, as it is a complicated function of energy, azimuth, and elevation direction. As seen in Figure \ref{fig:n_and_T_vs_Gfactor}, efficiency variations by this amount can slightly affect the estimated moments by $\sim$1\%. \par

\begin{figure*}[ht!] 
\scalebox{1.2}{\plotone{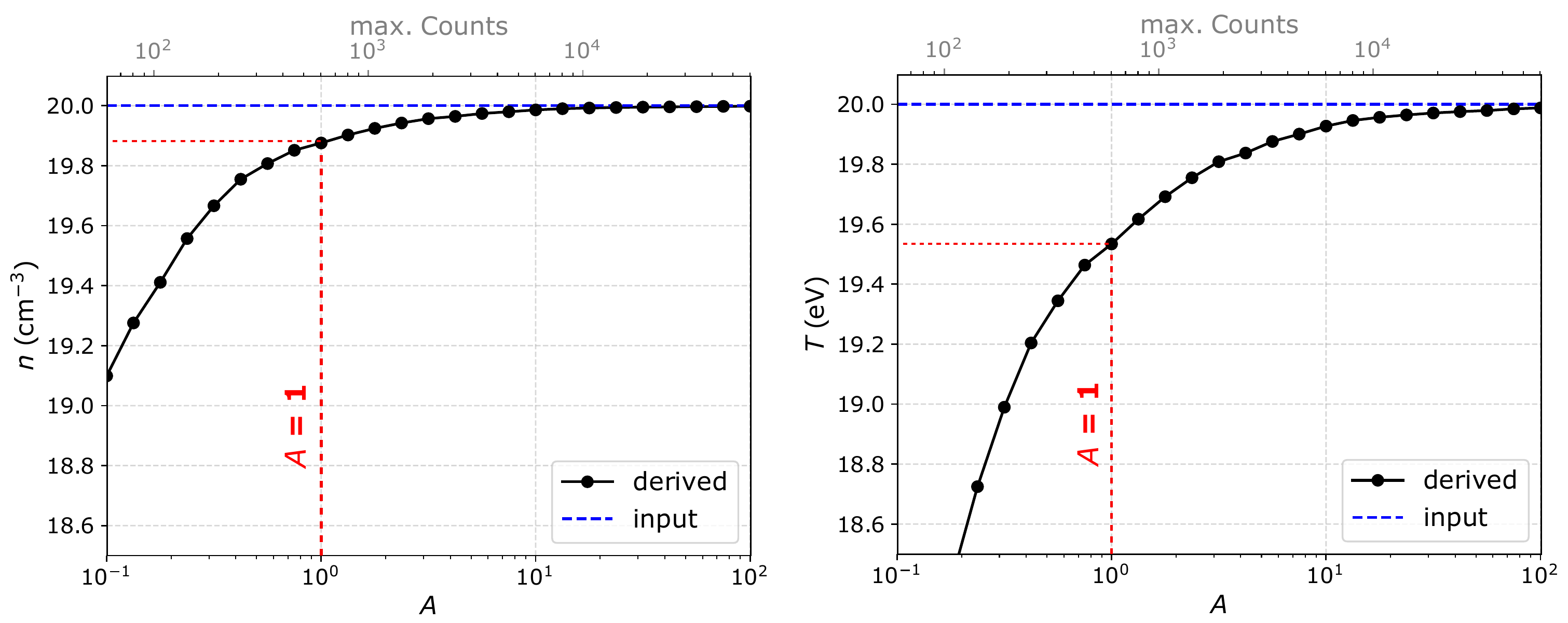}}
\caption{\added{The derived (left) density and (right) temperature as functions of the instrument's efficiency expressed in terms of the efficiency multiplier $A$, for solar wind in the absence of turbulence with $n_{\mathrm{0}}$=20 cm$^{-3}$, $T_{\mathrm{0}}$=20 eV, and $u_{\mathrm{0}}$=500 km s$^{-1}$. The blue dashed lines indicate the values of the input parameters, and the red dashed indicate the efficiency we use in all of our model calculations for the turbulent solar wind and the corresponding derived parameters. If $A$ is small, the distribution function is not fully resolved, therefore the derived $n$ and $T$ are underestimated. The derived parameters approach asymptotically the actual plasma parameters as $A$ increases (see the text for more).}}
\label{fig:n_and_T_vs_Gfactor}
\end{figure*}

The velocity distribution function of a turbulent plasma is fluctuating in velocity space. On the other hand, the F.O.V. and energy range of the instrument are finite and cannot capture the entire velocity space.} If the VDF is broader than or not entirely inside the instrument's F.O.V., the calculated moments are systematically underestimated. \added{In Figure \ref{fig:Alfven_Bulk} for example, we show the distribution functions of a plasma with $n_{\mathrm{0}}$ = 20 cm$^{-3}$, $T_{\mathrm{0}}$ = 20 eV and three different background bulk velocities: $u_{\mathrm{0}}$= 300, 500, and 800 km s$^{-1}$, respectively. For simplicity we set $\vec{u}_{\mathrm{0}}$ direction along the center of the F.O.V (x-direction). We set $V_{\mathrm{A}}\sim50$ km$\,$s$^{-1}$ and we let the bulk $u_{\mathrm{z}}$ component to fluctuate between -$V_{\mathrm{A}}$ and +$V_{\mathrm{A}}$. As the ratio $V_{\mathrm{A}}/u_{\mathrm{0}}$ increases, parts of the distribution function extend beyond the instrument's sampling range, causing an underestimation of the calculated moments. The underestimation of the moments is magnified as the distribution gets broader, which is the case for larger $T_{\mathrm{0}}/u_{\mathrm{0}}$. In addition, the instrument's absolute energy resolution decreases with energy, keeping $\Delta E / E$ = constant. Therefore, the instrument's ability to resolve the fluctuations decreases with increasing $u_{\mathrm{0}}$. We conclude that} \replaced{T}{t}he magnitude of the systematic errors varies with the plasma bulk parameters, but a detailed quantification of this effect is beyond the scope of this paper and the subject of a future study.

A detailed characterization of specific future data-sets should adopt our methods by adjusting our turbulence model to the specific plasma conditions. \added{We anticipate that the accuracy of the plasma moments depends on the plasma background parameters in a rather complicated way. Not only the amplitude and polarization properties of the fluctuations may change with changing plasma parameters in the solar wind, but also our instrument's ability to resolve them depends on the plasma parameters such as density, temperature, etc..\par 

Further, our turbulence model can be extended to include additional types of fluctuations such as fast-modes, whistler modes, coherent and pressure-balanced structures as discussed by \citet{Lacombe2014,Klein2012, Roberts2017}}. Such an extension would resemble the expected nature of the fluctuations more accurately; however, a detailed study of this type is beyond the scope of our work.  Advanced modeling of the plasma observations, could address additional sources of error which contribute to the total error in the derived plasma parameters. For instance, we note the contribution of the statistical counting error in any plasma measurements. According to counting statistics, every recorded number of particles $C$ has uncertainly $\delta C$ = $\sqrt{C}$ \citep[e.g.,][]{Livi2014, Nicolaou2014, Nicolaou2015a, Nicolaou2015b, Nicolaou2018, Elliott2016, Wilson2008, Wilson2012a, Wilson2012b, Wilson2015}. The relative statistical error $1/\sqrt{C}$ increases with decreasing counts, and could potentially propagate significant errors in the derived moments. As a rule of thumb, the statistical error increases with decreasing plasma flux through the instrument's aperture. Therefore, we expect larger statistical errors at larger heliocentric distances where the average plasma density is lower. A detailed characterization of the statistical error in SWA-PAS measurements is an ongoing work, which we will combine with the findings of this paper in order to characterize the future observations. 

\acknowledgments
G.N. \& C.J.O. are supported by the STFC Consolidated Grants to UCL/MSSL, ST/N000722/1 and ST/S000240/1. D.V. is supported by STFC Ernest Rutherford Fellowship ST/9003826/1. R.T.W is supported by the STFC Consolidated Grant to UCL/MSSL, ST/S000240/1. The authors thank Andrey Fedorov, Ali Varsani and Dhiren Kataria for helpful discussions.

%%%%%%%%%
%% APPENDIX %%
%%%%%%%%%

\appendix

           %%%%%%%%%%%%%%%%%
           %%%  METH: TURBULENCE   %%%
           %%%%%%%%%%%%%%%%%
\section{Model of turbulence spectrum} \label{sec:meth_turb}

While the analysis in Section \ref{sec:res} is performed in the spacecraft frame $(x^{\prime},y^{\prime},z^{\prime})$, we now adopt a coordinate system $(x,y,z)$ in which the background magnetic field $\vec{B_0}$ is parallel to $\hat{z}$. Both reference frames are connected through a rotation around the common $y^{\prime}/y$ axis. We define the background density $n_0$, background temperature $T_0$ and background velocity $\vec{u}_0$. We model plasma turbulence through a superposition of Alfv\'{e}nic ($\Delta B_{\mathrm{A}}$) and slow-mode ($\Delta B_{\mathrm{S}}$) fluctuations:

\begin{equation} \label{eq:superposition}
\Delta \vec{B}(t)= \Delta \vec{B}_{\mathrm{A}}(t) + \Delta \vec{B}_{\mathrm{S}}(t)  =C_1 \sum_i{\delta \vec{B}_i}(t)+C_2 \sum_j{\delta \vec{B}_j}(t),
\end{equation}
\noindent where $C_1$ and $C_2$ are normalization constants.

Similarly, the plasma density fluctuations are
\begin{equation} \label{eq:density_superposition}
\Delta n(t)= \Delta n_{\mathrm{A}}(t) + \Delta n_{\mathrm{S}}(t),
\end{equation}

\noindent and the components of the velocity fluctuations $\Delta \vec{u}(t)$ are
\begin{equation} \label{eq:upar_superposition}
\Delta u_{\parallel}(t)= \Delta u_{\parallel \mathrm{A}}(t) + \Delta u_{\parallel \mathrm{S}}(t)
\end{equation}
\noindent and
\begin{equation} \label{eq:uperp_superposition}
\Delta \vec{u}_{\perp}(t)= \Delta \vec{u}_{\perp \mathrm{A}}(t) + \Delta \vec{u}_{\perp \mathrm{S}}(t).
\end{equation}

The magnetic field and plasma fluctuations are convected over the spacecraft and thus only functions of time $t$. In our model, we consider the plasma particles to follow a Maxwell distribution function with changing bulk parameters:

\begin{equation} \label{eq:Maxwellian}
f(\vec{u},t)=(n_0+\Delta n(t)) \left( \frac{m}{2 \pi k_{\mathrm{B}} T_0} \right) ^{3/2} \exp{ \left ( -\frac{m[\vec{u}-(\vec{u}_0+\Delta{\vec{u}(t)})]^2}{2 k_{\mathrm{B}} T_0} \right )},
\end{equation}
\noindent where $k_{\mathrm{B}}$ is the Boltzmann constant. 

In the following subsections, we describe in detail the simulation setup for the Alfv\'en and slow-mode waves.

%%%%%%%%%%%
%% Alfven waves %%
%%%%%%%%%%%
\subsection {Alfv\'en-wave spectrum} \label{sub_sec:Alfven_waves_spectrum}
For each Alfv\'en wave harmonic, we assume
\begin{equation} \label{eq:harm}
{\delta \vec{B}_i}=\vec{A}_i \mathrm{cos}(\vec{k}_i\cdotp \vec{u}_0 t +\Psi_i),
\end{equation}
\noindent where $\vec{A}$ is the amplitude, $\vec{k}$ the wave vector, and $\Psi_i$ the phase. In Equation (\ref{eq:harm}), we adopt Taylor's hypothesis, assuming that the observed fluctuations are due to the convection of frozen turbulence in the mean flow of the solar wind with velocity $\vec{u}_0$.  

\noindent Each harmonic $\delta \vec{B}_i$ has a wave vector $\vec{k}$ with components:

\begin{eqnarray} \label{eq:k_comp}
k_{x} &=& k\,\mathrm{sin}\theta_k \mathrm{cos} \phi_k, \nonumber \\
k_{y} &=& k\,\mathrm{sin}\theta_k \mathrm{sin}\phi_k, \nonumber\\
k_{z} &=& k\,\mathrm{cos}{\theta_k},
\end{eqnarray}

\noindent where $\theta_k$ is the angle between $\vec{k}$ and $\vec{B}$ and $\phi_k$ the azimuthal angle of $\vec{k}$. We define the components of $\vec{k}$ along and perpendicular to $\vec{B}$ as $k_{\parallel}$ and $k_{\perp}$respectively. Then, according to Equation (\ref{eq:k_comp}),

\begin{eqnarray} \label{eq:k_par_perp}
k_{\parallel} &=& k\,\mathrm{cos}{\theta_k}, \nonumber \\ 
k_{\perp} &=& k\,\mathrm{sin}\theta_k.
\end{eqnarray}

\noindent In our model, we consider the superposition of waves with a $k_{\perp}$ component with:

\begin{equation} \label{eq:k_perp_range}
10^{-4} \leq k_{\perp}\rho_{\mathrm{g}} < 10^3,
\end{equation}

\noindent where $\rho_{\mathrm{g}}$ the proton gyroradius. We model fluctuations with $k_{\perp}\rho_{\mathrm{g}}$ $\leq$ 1 as Alfv\'en waves (AWs), and those with $k_{\perp}\rho_{\mathrm{g}}$ $>$ 1 as kinetic Alfv\'en waves (kAWs). We discretize our spectrum in 71 $k_{\perp}$ steps; 41 steps within the range of AWs and 30 within the range of kAWs. As in \citet{Chandran2010}, for the amplitude of each harmonic, we set

\begin{equation} \label{eq:Amp}
|\vec{A}|=10^{-4/3}B_0(k_{\perp}\rho_{\mathrm{g}})^{-\gamma_s},
\end{equation}

\noindent where the spectral index

\begin{equation} \label{eq:sp_index}
  \gamma_s=\left\{
    \begin{array}{l}
      1/3 \quad \text{for} \quad k_{\perp}\rho_{\mathrm{g}}\leq1, \\
      2/3 \quad \text{for} \quad k_{\perp}\rho_{\mathrm{g}}>1.
    \end{array}
  \right.
\end{equation}

The parallel component of the wave vector is

\begin{equation} \label{eq:k_par}
k_{\parallel}\rho_{\mathrm{g}}=10^{-4/3}( k_{\perp}\rho_{\mathrm{g}})^{(1-\gamma_s)},
\end{equation}

\noindent according to the critical-balance assumption \citep{Goldreich1995}. The spectrum is continuous at $k_{\perp}\rho_{\mathrm{g}}=1$, and $\vec{A} \perp \vec{B_0}$ and $\vec{A} \perp \vec{k}$. Equations (\ref{eq:sp_index}) and (\ref{eq:k_par}) guarantee that, in the low-frequency limit, the turbulence is isotropic ($k_{\perp}\rho_{\mathrm{g}}=k_{\parallel}\rho_{\mathrm{g}}=10^{-4}$) and becomes highly anisotropic with increasing frequency \citep[e.g.,][]{Horbury2012,Chen2016}.

For each of the 71 values of $|\vec{k}|$, we set 101 $\phi_k$ angle values, reaching from 0 to 2$\pi$ in uniform bins. In addition, for each combination of $|\vec{k}|$ and $\phi_k$, we include one wave propagating in the +$k_{\parallel}$-direction and one wave propagating in the -$k_{\parallel}$-direction. All $71 \times 101 \times 2=14\,342$ waves that construct the spectrum have a different phase, randomly selected from the range from 0 to 2$\pi$.
After some algebra, combining Equations (\ref{eq:harm}) through (\ref{eq:Amp}), the sum in Equation (\ref{eq:superposition}) can be expressed as

\begin{eqnarray} \label{eq:dBx}
&&\Delta B_{\mathrm{A},x}=C_1 \sum_{l=1}^{71} \sum_{m=1}^{101} \sum_{n=1}^{2} -10^{-4/3}B_0(k_{\perp,l}\rho_{\mathrm{g}})^{-\gamma_s} \mathrm{sin}\phi_{k,m} \nonumber \\
&&\times \, \mathrm{cos}[ (k_{\perp,l}\mathrm{cos}\phi_{k,m} \vec{u}_0 \cdot \hat{x} + k_{\perp,l}\mathrm{sin}\phi_{k,m} \vec{u}_0 \cdot \hat{y} \nonumber \\
&&+ \, (-1)^{n}k_{\parallel} \vec{u}_0 \cdot \hat{z}) \, t + \Psi_{lmn} ],
\end{eqnarray}
\noindent and

\begin{eqnarray} \label{eq:dBy}
&&\Delta B_{\mathrm{A},y}=C_1 \sum_{l=1}^{71} \sum_{m=1}^{101} \sum_{n=1}^{2} 10^{-4/3}B_0(k_{\perp,l}\rho_{\mathrm{g}})^{-\gamma_s} \mathrm{cos} \phi_{k,m} \nonumber \\
&&\times \, \mathrm{cos}[ (k_{\perp,l}\mathrm{cos}\phi_{k,m} \vec{u}_0 \cdot \hat{x} + k_{\perp,l} \mathrm{sin}\phi_{k,m} \vec{u}_0 \cdot \hat{y} \nonumber \\
&&+ \, (-1)^{n}k_{\parallel} \vec{u}_0 \cdot \hat{z}) \, t + \Psi_{lmn} ].
\end{eqnarray}

The density and velocity fluctuations are modeled as

\begin{equation} \label{eq:dn}
\frac{\Delta n_{\mathrm{A}}(t)}{n_0}=C_1 \sum_{l=1}^{71}\sum_{m=1}^{101}\sum_{n=1}^{2}{ \xi_{lmn} \frac{\delta B_{\perp,lmn}(t)}{B_0}    },
\end{equation}

\begin{equation} \label{eq:dupar}
\frac{\Delta u_{\parallel \mathrm{A}}(t)}{V_{\mathrm{A}}}=C_1 \sum_{l=1}^{71}\sum_{m=1}^{101}\sum_{n=1}^{2}{\chi_{\parallel,lmn} \frac{\delta B_{\perp,lmn}(t)}{B_0} },
\end{equation}

\noindent and
\begin{equation} \label{eq:duperp}
\frac{\Delta \vec{u}_{\perp \mathrm{A}}(t)}{V_{\mathrm{A}}}=C_1 \sum_{l=1}^{71}\sum_{m=1}^{101}\sum_{n=1}^{2}{ \chi_{\perp,lmn} \frac{\delta \vec{B}_{\perp,lmn}(t)}{B_0} },
\end{equation}

\noindent for each harmonic $\delta \vec{B}$ of the spectrum, according to the two-fluid solutions for $\xi_{lmn}$, $\chi_{\parallel,lmn}$, and $\chi_{\perp,lmn}$ by \citet{Hollweg1999} and \citet{Wu2019}. In Equations (\ref{eq:dupar}) and (\ref{eq:duperp}), $V_{\mathrm{A}}$ is the Alfv\'en speed.

The normalization constant $C_1$ is chosen so that the root-mean-square (rms) value of the magnetic field fluctuations over a large interval $\Delta T$ is equal to the background magnetic field,

\begin{equation} \label{eq:rms1}
C_1=\frac{B_0}{\sqrt{\displaystyle{\frac{1}{\Delta T}\sum_{t=0}^{t=\Delta T}{|\Delta \vec{B}_A(t)|^2 }}}}.
\end{equation} 
\noindent For Equation \ref{eq:rms1}, we use $\Delta T \, \sim10^{5}$ s, which is $\sim$10 times larger than the period of the isotropic fluctuations with $k_{\perp}\rho_{\mathrm{g}}$  = 10$^{-4}$.     
%%%%%%%%%%%%%
%% slow mode waves %%
%%%%%%%%%%%%%
\subsection{Slow mode spectrum} \label{sub_sec:slow_mode_spectrum}

The harmonics of the slow-modes follow the same form as Equation (\ref{eq:harm}). In our model, we simulate slow modes with:
\begin{equation} \label{eq:k_perp_range_sm}
10^{-4} \leq k_{\perp}\rho_{\mathrm{g}} \leq 1.
\end{equation}

Equations (\ref{eq:Amp}) through (\ref{eq:k_par}) also apply to the slow modes in our model. The dimensionless factors $\xi_{lmn}$ and $\chi_{\parallel,lmn}$ are calculated as by \citet{Verscharen2016} and \citet {Verscharen2017}. The normalization constant $C_{2}$ is chosen so that the total amplitude of compressive sow-mode-like fluctuations is 10 \% of the total amplitude of the incompressive Alf\'enic fluctuations: 

\begin{equation} \label{eq:rms2}
C_2=\frac{0.1 B_0}{\sqrt{\displaystyle{\frac{1}{\Delta T}\sum_{t=0}^{t=\Delta T}{|\Delta \vec{B}_S(t)|^2 }}}},
\end{equation} 

\begin{figure*}[ht!] 
\scalebox{1.2}{\plotone{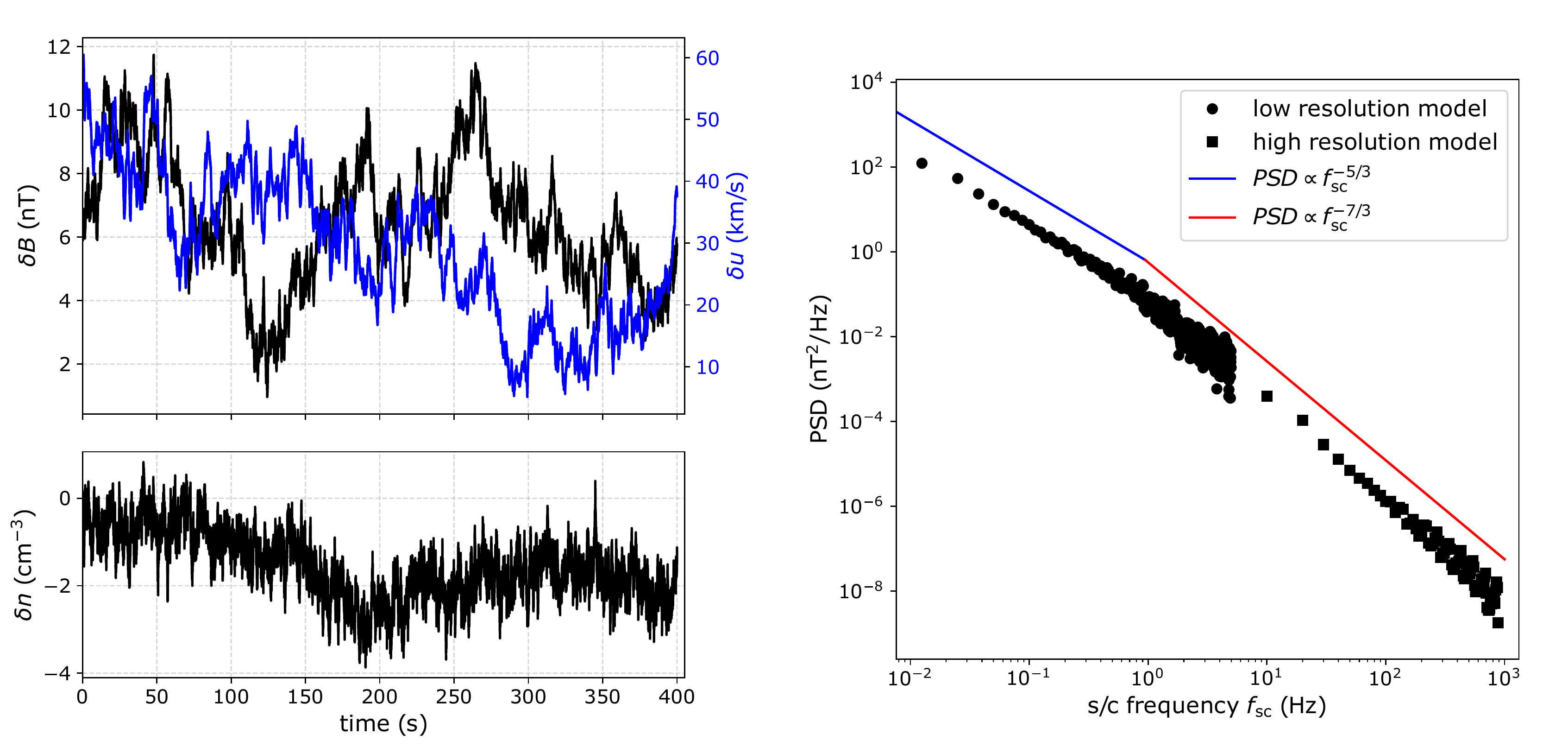}}
\caption{(Top left) time series of the modeled magnetic field (black) and bulk speed (blue) fluctuations for a turbulent plasma in our model with background parameters $n_0$ = 20 cm$^{-3}$, $T_0$ = 20 eV, $B_0$ = 10 nT, and $u_0$ = 500 km\,s$^{-1}$. (Bottom left) time series of the modeled density fluctuations. (Right) the power spectral density of magnetic-field fluctuations with combined low resolution (10$^{-1}\,\mathrm{s}$, bullets) and high resolution (10$^{-4}\,\mathrm{s}$, squares) model data series. We overplot $\mathrm{PSD} \propto f_{\mathrm{sc}}^{-5/3}$ (blue) and $\mathrm{PSD} \propto f_{\mathrm{sc}}^{-7/3}$ (red) for reference.}
\label{fig:dB_du_complete}
\end{figure*}

%% total spectrum settings
We model density and velocity fluctuations to construct the velocity distribution function using Equation (\ref{eq:Maxwellian}) for $B_0=$10 nT, $u_0$=500 km\,s$^{-1}$, $T_{0}$=20 eV, and $n_{0}$=20 cm$^{-3}$. We set the bulk velocity along the spacecraft's $\hat{x}$ direction (anti-sunward along Sun-spacecraft line) and the magnetic field vector $45^{\circ}$ elevated in the $x-y$ (top hat) plane. We model time series with a resolution $10^{-4}\, \mathrm{s}$, which is 10 times shorter than the SWA-PAS acquisition time for one energy and one elevation direction. We also model time series with a lower resolution ($10^{-1}\, \mathrm{s}$), which we use only to examine the modeled spectrum in the lower frequency domain. The top left panel of Figure \ref{fig:dB_du_complete} shows a time series of 400~s of the high resolution modeled magnetic field and proton speed fluctuations. The bottom left panel shows a time series of the density fluctuations for the same time interval, while the panel on the right shows the power spectral density of the magnetic field fluctuations, combining both the $10^{-4}\,\mathrm{s}$ and the $10^{-1}\, \mathrm{s}$ resolution models. The spectral density follows the expected $f_{\mathrm{sc}}^{-5/3}$ and $f_{\mathrm{sc}}^{-7/3}$ profiles. Note that, half of the harmonics in our spectrum, propagate along the magnetic field, while the other half propagate in the opposite direction. Therefore, we do not observe any consistent correlation or anti-correlation between the magnetic field and the plasma fluctuations in Figure \ref{fig:dB_du_complete}. \added{Test studies of imbalanced turbulence using the extreme cases in which all waves propagate (i) parallel and (ii) anti-parallel to $\vec B_0$ (not shown here) lead to very similar observations to the ones shown for the balanced case in this work. Despite the fact that imbalanced turbulence exhibits persistent averaged correlations or anti-correlations between $\vec B$ and $\vec u$, the virtually identical spectra compared to the one we use (Figure \ref{fig:dB_du_complete}) lead to almost identical results.}

\end{document}